\documentclass{article}
\usepackage{spconf,amsmath,graphicx}
\usepackage{booktabs}

\title{End-to-end Streaming model for Low-Latency Speech Anonymization}
\name{Waris Quamer, Ricardo Gutierrez-Osuna}
\address{Department of Computer Science and Engineering, Texas A\&M University, United States\\
\{quamer.waris, rgutier\}@tamu.edu}
\begin{document}
%\ninept
%
\maketitle
\begin{abstract}
Speaker anonymization aims to conceal cues to speaker identity while preserving linguistic content. Current machine learning based approaches require substantial computational resources, hindering real-time streaming applications. To address these concerns, we propose a streaming model that achieves speaker anonymization with low latency. The system is trained in an end-to-end autoencoder fashion using a lightweight content encoder that extracts HuBERT-like information, a pretrained speaker encoder that extract speaker identity, and a variance encoder that injects pitch and energy information. These three disentangled representations are fed to a decoder that re-synthesizes the speech signal. We present evaluation results from two implementations of our system, a full model that achieves a latency of $230 ms$, and a lite version (0.1x in size) that further reduces latency to $66 ms$ while maintaining state-of-the-art performance in naturalness, intelligibility, and privacy preservation.
\end{abstract}
\section{Introduction}
\label{sec:intro}

The task of speaker anonymization is to transform utterances to hide the identity of the speaker (while preserving their linguistic content). Speaker anonymization provides privacy protection and confidentiality in a range of applications, including customer service interactions, voice-operated  virtual assistants, legal proceedings, and medical consultations. Moreover, speaker anonymization addresses ethical and responsible use of speech data, aligning with privacy regulations and safeguarding individuals' rights.

Existing machine learning (ML) based approaches to speaker anonymization follow a cascaded automatic speech recognition (ASR) -- text-to-speech (TTS) architecture \cite{meyer2023anonymizing, pierre22_interspeech}. An ASR module produces a text transcription that is speaker independent but eliminates emotional cues that may otherwise be of use for downstream applications. Moreover, existing systems for speaker anonymization are computationally heavy, operate in a non-streaming fashion, and/or have high latency on CPU devices as opposed to GPUs. For speech anonymization to be used in the field, it must operate at real time (or faster), exhibit low latency, require minimal future context and be compatible with low-resource devices  (e.g., smartphones). 

To address these needs, we propose an end-to-end streaming model suitable for low-latency speaker anonymization.  Our model draws inspiration from neural audio codecs \cite{zeghidour2021soundstream, defossez2022high} for audio compression in low-resource streaming settings. Our key strategy that enables streaming is to replace traditional non-causal computationally intensive networks (\textit{e.g.}, ASR or self-supervised learning based models) for encoding linguistic content with a lightweight convolutional neural network (CNN) based architecture. Our proposed architecture consists of: (a) a streaming waveform encoder that generates a speaker-independent content representation from waveforms, (b) a pseudo-speaker generator that produces an anonymized speaker representation (i.e., an embedding) from the input speech, (c) a speaker/variance adapter that adds speaker, pitch and energy information to the content representation, and (d) a streaming decoder that consumes the speaker/variance adapted linguistic representation and the corresponding speaker embedding to generate the final anonymized audio waveform. Our system is trained in an auto-encoder fashion, which reconstructs the input \textit{conditioned} on the speaker embeddings generated using pre-trained speaker encoders \cite{desplanques2020ecapa, snyder2018x}. During inference, a pseudo-speaker generator produces a target speaker embedding with cosine distance greater than 0.3 from the source embedding, ensuring that the re-synthesized utterance sounds as if a different (i.e., anonymized) speaker had produced it. Additionally, the speaker/variance adapter is used to modulate pitch and energy values to further enhance privacy and control the similarity of the synthesized speech with the source audio. We show that our lightweight convolutional neural network (CNN) based architecture achieves similar performance as traditional content encoders.

We perform experiments on two versions of our model, a \textit{Base} version that can perform real-time streaming synthesis with a latency of $230 ms$ and a \textit{Lite} version (having 0.1x the number of parameters) that further reduces latency to $66 ms$ while maintaining state-of-the-art performance on naturalness, intelligibility, privacy and speaker identity transfer\footnote{https://warisqr007.github.io/demos/stream-anonymization/}. 

\section{Related Work}
\label{sec:relatedwork}

\subsection{Voice conversion}
Speaker anonymization is closely related to voice conversion (VC). However, whereas VC seeks to transform utterances from a source speaker to match the identity of a (known) target speaker, speaker anonymization only requires that the transformed speech be sufficiently different from the source speaker to conceal their identity. 

The first step in conventional VC architectures is to disentangle the linguistic content of an utterance from speaker-specific attributes. As an example, cascaded ASR-TTS architectures \cite{huang2020sequence} use an ASR model to transcribe the input utterance into text, followed by a TTS model that converts the text back into speech --conditioned on a speaker embedding. Variants of this approach replace the ASR module with acoustic models that generate a more fine-grained representation than text, such as phonetic posteriorgrams (PPGs) \cite{liu2021any}.  Recent approaches have also used information bottlenecks to disentangle linguistic content form speaker identity \cite{chan2022speechsplit2}. A major drawbacks of the latter approach is that information bottlenecks must be carefully designed and are sensitive to the dimension of latent space. Other techniques include instance normalization \cite{chen2021again}, use of mutual information loss \cite{wang21n_interspeech}, vector quantization \cite{wang21n_interspeech, quamer23_interspeech}, and adversarial training \cite{ding2020improving}. To enable streaming, recent VC methods use a streaming ASR to extract PPGs \cite{tran22_interspeech} or streaming ASR sub-encoders \cite{yang2022streamable, chen2023streaming} to generate linguistic content, and then perform VC through causal architectures that require limited future contexts.

\subsection{Speaker anonymization}
Speaker anonymization approaches can be broadly divided into two categories: digital signal processing (DSP) and machine learning (ML) based. DSP methods include formant-shifting using McAdams coefficients \cite{patino21_interspeech}, frequency warping \cite{qian2019speech}, or a series of steps consisting of vocal tract length normalization, McAdams transformation and modulation spectrum smoothing \cite{kai2022lightweight}. Additionally, modifications to pitch \cite{tavi2022improving} and speaking rate \cite{dubagunta2022adjustable} are used. DSP models are significantly smaller (i.e., fewer parameters) than ML models, which results in efficient and speedy execution. However, the types of global transforms used in DSP methods cannot fully remove speaker-dependent cues, making them vulnerable to ML-based speaker verification systems  \cite{meyer2023anonymizing}.

ML methods for speaker anonymization follow the conventional VC framework of disentangling linguistic content from speaker identity, but then replace the latter with a speaker embedding that is different (anonymized) from the source. Various methods have been proposed to select this anonymized speaker embedding. For example, Srivastava \textit{et al.} \cite{srivastava20_interspeech} generate anonymized embeddings by randomly selecting N speaker vectors from a pool of speakers farthest from the source, using e.g., cosine distance, whereas Perero-Codosero \textit{et al.} \cite{perero2022x} use an autoencoder architecture with an adversarial training module that removes speaker, gender, and accent information. 
% Alternatively, Turner \textit{et al.} \cite{turner2022generating} proposed a Gaussian mixture model to sample the anonymized speaker representation within a principal component analysis (PCA) reduced space, and thus maintaining the distributional properties of the original speaker embeddings. 
Other approaches use look-up tables \cite{yao22_spsc} or generative adversarial networks \cite{meyer2023anonymizing}  to generate pseudo-speakers. Our approach follows the latter: we combines a GAN-based pseudo-speaker generator with a  streaming model to enable real-time speaker anonymization with low latency.

\section{Method}
\label{sec:method}
\begin{figure}
    \centering
    \includegraphics[scale=0.47]{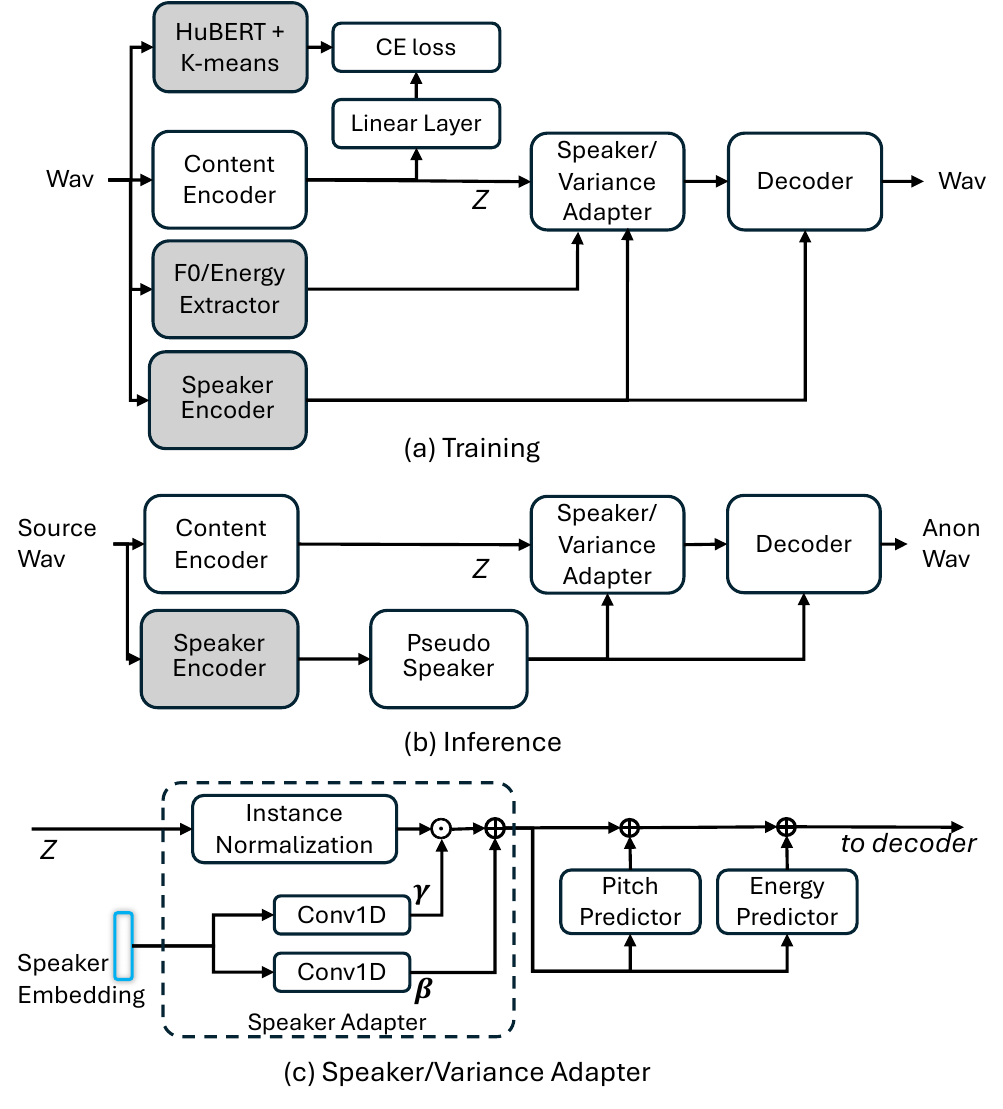}
    \caption{Block diagram of the proposed anonymization system. (a) training workflow (b) inference workflow (c) speaker/variance adapter.}
    \label{fig:block_diagram}
\end{figure}

The proposed system is illustrated in Figure \ref{fig:block_diagram}. Anonymization takes place in two steps, (1) generating a fixed (i.e., off-line) anonymized speaker embedding personalized to the source speaker and, (2)  using this fixed anonymized speaker embedding and the streaming speech synthesizer to synthesize anonymized speech that only preserves the linguistic content of the source speech. To generate the anonymized embedding, a reference waveform from the source speaker is passed to the pre-trained speaker encoder, which then produces the source speaker embedding. The pseudo speaker generator receives this source speaker embedding and generates the anonymized version (see Figure \ref{fig:block_diagram}b). To synthesise speech signals, the content encoder receives streaming chunks of waveform and converts it into a hidden representation $z$ that contains the linguistic content disentangled from the speaker representation. The content information $z$ and the anonymized speaker embedding (generated in the previous step) is passed to the speaker/variance adapter. The speaker/variance adapter, first, conditions the anonymized speaker embedding on the content representation $z$ and then adds pitch and energy values. The decoder receives the output of the speaker/variance adapter and the anonymized speaker embedding to generate the final anonymized waveform.

We train two versions of our proposed system, a \textit{base} and a \textit{lite} version. Below, we describe each component of our system and the training procedure in detail.

\subsection{Content encoder}
% Our encoder uses an acoustic model  to generate a  speaker-independent linguistic representation. The acoustic model generates a phonetic posteriorgram (posterior probability that each speech frame belongs to a predefined set of phonetic units). We take the output of the last hidden layer as bottleneck features (BNFs) which contain similar phonetic information. Instead in our study, following \cite{van2022comparison}, 
The content encoder consumes the wav signal to predict discrete speech units produced by discretizing the output speech representation from a pretrained HuBERT model \cite{hsu2021hubert} into one of N codewords or pseudo-labels \cite{van2022comparison}. Our content encoder architecture follows that of HiFiGAN \cite{kong2020hifi}, except all transposed convolutions in HiFiGAN are replaced with strided causal convolutions to downsample the input waveform. Additionally, to support streaming applications, we replace all vanilla CNN layers in HiFiGAN with causal CNNs so that the prediction only considers the past context and does not rely on future audio frames. For both versions of our model (base and lite), we use downsampling rates of $[2, 2, 4, 4, 5]$. The residual blocks have kernel sizes as $[3, 7, 11]$ with dilation rates as $[[[1, 1], [3, 1], [5, 1]] * 3]$ (please refer \cite{kong2020hifi} for details). The difference between \textit{base} and the \textit{lite} version is the dimension of the hidden representation $z$ (the output of the encoder): 512 dimensions for the \textit{base} version and 128 dimensions for the \textit{lite} version.

\subsection{Speaker encoder and pseudo-speaker generator}
Speaker verification or classification systems generally use speaker embeddings to represent the characteristics or timbre of a speaker's voice. Widely used speaker encoders include the GE2E model \cite{wan2018generalized}, X-vectors \cite{snyder2018x} and ECAPA-TDNN \cite{desplanques2020ecapa}. Our system concatenates embeddings generated from X-vectors and ECAPA-TDNN models, since these two models have been shown to complement each other \cite{meyer2023anonymizing}.

To perform speaker anonymization, we use a pseudo-speaker generator that takes the original speaker embedding as input and outputs an artificially generated speaker embedding such that the generated anonymized speaker embedding has a cosine distance greater than 0.3 as compared with the original speaker embedding. Our pseudo-speaker generator follows a GAN-based architecture \cite{meyer2023anonymizing} and is trained separately. The generator is trained to receive a random vector sampled from a standard normal distribution $N (0, 1)$ as input and output a vector of the same shape as the original speaker embedding. The discriminator is trained to discriminate w.r.t the quadratic Wasserstein distance and transport cost \cite{liu2019wasserstein} between the artificial and the original speaker embeddings. 

\subsection{Speaker/Variance adapter}
The speaker/variance adapter aims to add speaker, pitch, and energy (i.e., variance) information to the speaker-independent content representation and provides a way to control them \cite{ren2020fastspeech}. The speaker/variance adapter consists of three modules: (a) speaker adapter, (b) pitch predictor, and (c) energy predictor (see Figure \ref{fig:block_diagram}c).

The speaker adapter conditions the speaker embedding on the content representation $z$, and passes it to the pitch and energy predictors. The speaker adapter is based on adaptive instance normalization (adaIN) \cite{huang2017arbitrary} and feature-wise linear modulation (FiLM) \cite{perez2018film}. The conditioning goes as follows. First, we apply instance normalization to the input feature representation, and then transform it with scale and bias parameters learned through two 1D CNNs that take speaker embeddings as input. The use of instance normalization was motivated by a prior work \cite{chen2021again} that showed instance normalization being helpful in removing residual speaker information.
 
Pitch and energy predictor estimate pitch and energy values based on speaker adapted content representation $z$. During training, we use the ground-truth pitch and energy values to train the pitch and energy predictors. At inference, the output of the pitch and energy predictor are added to the speaker adapted $z$. The pitch and energy predictors have similar architecture consisting of a 2-layered 1D causal CNNs (kernal size 3) with ReLU activation, followed by layer normalization and dropout layer and an additional 1D CNN (with kernel size 1) to project pitch and energy values on the latent representation.

\subsection{Decoder}
The decoder follows the same design and training procedure as HiFiGAN \cite{kong2020hifi} and can be seen as a mirror-image of the content encoder. Similar to the content encoder, all vanilla CNNs are replaced with causal CNNs. The decoder receives speaker/variance adapted latent representation along with speaker embedding and directly generates waveform signal without any intermediate mel-spectrogram generation. We additionally adapt the output of each residual block of the decoder to the speaker embedding. In our experiments, we observed that doing so gave better speaker transfer performance. For both versions of $base$ and $lite$ version of our model, we use upsampling rates of $[5, 4, 4, 2, 2]$. The residual blocks have kernel sizes as $[3, 7, 11]$ with dilation rates as $[[[1, 1], [3, 1], [5, 1]] * 3]$.

\subsection{Training}
The training workflow is described in Figure \ref{fig:block_diagram}a. The proposed system is trained end-to-end similar to an autoencoder, reconstucting the same waveform at output that fed as input. The content encoder is trained to predict the pseudo-labels generated through a HuBERT/Kmean module using cross-entropy loss. The pitch and energy predictor in the variance adapter apply mean-squared error loss for pitch and energy prediction. At the output of the decoder, Following the HiFiGAN architecture \cite{kong2020hifi}, we apply a combination of adversarial losses at the output of the decoder, including feature loss, multi-period discriminator loss, multi-scale discriminator loss, multi-resolution STFT loss \cite{yamamoto2020parallel} and mel-spectrogram reconstruction loss. These discriminators have similar architecture as those in HiFiGAN, including weighting schemes to compute the total decoder loss. The final training loss is the summation of content loss, pitch/energy error loss and decoder losses. We apply a stop-gradient operation to prevent gradient flow (i.e., back-propagation) from the decoder to the encoder, to ensure that the speaker information is not leaked via the content representation. This operation effectively decouples it from the rest of the system; in other words, the encoder and the rest of the system can potentially be trained sequentially as two independent modules.

\section{Experimental setup}
\label{sec:setup}
We trained our system on the LibriTTS corpus \cite{zen2019libritts} following guidelines for the Voice Privacy Challenge 2022 (VPC22) \cite{tomashenko2022voiceprivacy}. All our experimental results are presented on the LibriTTS dev and test set, which were not part of the training. We use a pretrained HuBERT-\textit{base}\footnote{https://github.com/facebookresearch/fairseq} and extract the the output from its 9th layer. We set the number of cluster centroids to 200. For all our experiments, we use a sampling rate of 16 kHz and batch size of 16 with the AdamW optimizer with a learning rate of $2*10^{-4}$ annealed down to $10^{-5}$ by exponential scheduling. The encoder is first pretrained for 300k steps (for training stability), and then trained together with the decoder for an additional 800k steps. The pretrained speaker encoders were taken from speechbrain \cite{ravanelli2021speechbrain}. The pseudo speaker embedding generator follows the training procedure described in \cite{meyer2023anonymizing} and trained on VoxCeleb 1 and 2 \cite{Nagrani17, Chung18b}. All our models are trained using two NVIDIA Tesla V100 GPUs for approximately two weeks.

\section{Results}
\label{sec:results}
We evaluated our system on a series of subjective and objective measures of synthesis latency, synthesis quality, privacy as well as speaker transfer ability. We compare our results against five baselines: three state-of-the-art VC models (VQMIVC\footnote{https://github.com/Wendison/VQMIVC} \cite{wang21n_interspeech}, QuickVC\footnote{https://github.com/quickvc/QuickVC-VoiceConversion} \cite{guo2023quickvc}, and DiffVC\footnote{https://github.com/huawei-noah/Speech-Backbones/tree/main/DiffVC} \cite{popov2021diffusion}) and two speaker anonymization models\footnote{https://github.com/Voice-Privacy-Challenge/Voice-Privacy-Challenge-2024}, a DSP-based model \cite{patino21_interspeech} (baseline B2 from  VPC22) and a ML-based model (to which we refer as B3) that uses a transformer-based ASR and a Fastspeech2-based TTS with a WGAN-based anonymizer \cite{meyer2023anonymizing}. The five baseline models are trained on the same dataset as our proposed system, and we use their pretrained checkpoints obtained from their corresponding github repositories. We could not find any open-source streaming speech synthesis model and hence were unable to include them as baselines. We evaluate our model using the same Libri-TTS evaluation split as VPC22. For the VC baselines we randomly select a speaker from the CMU Arctic corpus \cite{kominek2004cmu} as the target speaker.

\subsection{Synthesis Latency}
Our pretrained HuBERT model produces speech frames at $50 Hz$, so the smallest chunk size that our model can process is $20 ms$. In this section, we present the synthesis latency and the real-time factors (RTF) for the base and lite versions for our model, for various chunk sizes between $20 ms$ and $140 ms$ on both CPU and GPU devices. Latency is defined as the sum of the chunk size and the average time the model takes to synthesize that chunk. RTF is the ratio of the system's average processing time to the chunk size. For a system to be real-time, the latency should be less than twice the chunk size, meaning RTF should be less than 1. Results are summarized in Table \ref{tab:sythesis_latency}. On GPU, our base version can operate in real-time for the chunk size of $40 ms$ with a latency of $64 ms$, while on CPU the base model can be real-time for chunk size of $120 ms$ with a latency of $230 ms$. In case of the lite version, the model is real-time for chunk size of $20 ms$ with a latency of $38 ms$ on GPU and can operate in real-time for $40 ms$ with latency of $66 ms$ on the CPU. For our test set of experiments we set the chunk size of $120 ms$ and $40 ms$ for the base and lite versions respectively.

\begin{table*}[hbt!]
\caption{Synthesis latency and real-time factors (RTF) on CPU and GPU devices.}
\label{tab:sythesis_latency}
\centering
\begin{tabular}{c cc cc cc cc}
\toprule
\textbf{Chunk Size} & \multicolumn{4}{c}{\begin{tabular}[c]{@{}c@{}c@{}c@{}}\textbf{CPU} (ms)\end{tabular}} & \multicolumn{4}{c}{\begin{tabular}[c]{@{}c@{}c@{}c@{}}\textbf{GPU} (ms)\end{tabular}} \\ 
\midrule

(ms) & \multicolumn{2}{c}{\textbf{base}} & \multicolumn{2}{c}{\textbf{lite}} & \multicolumn{2}{c}{\textbf{base}} & \multicolumn{2}{c}{\textbf{lite}} \\

     & \multicolumn{1}{c}{latency} & RTF & \multicolumn{1}{c}{latency} & RTF & \multicolumn{1}{c}{latency} & RTF & \multicolumn{1}{c}{latency} & RTF \\
\midrule

20 & \multicolumn{1}{c}{83.82}   &  3.19 & \multicolumn{1}{c}{43.01} & 1.15  & \multicolumn{1}{c}{44.61}   &  1.23 & \multicolumn{1}{c}{38.36} & 0.92 \\ 

40 & \multicolumn{1}{c}{115.23}  &  1.88 & \multicolumn{1}{c}{66.27} & 0.66  & \multicolumn{1}{c}{64.38}  &  0.61 & \multicolumn{1}{c}{58.70} & 0.47 \\ 

60 & \multicolumn{1}{c}{170.93}  &  1.85 & \multicolumn{1}{c}{88.03} & 0.47 & \multicolumn{1}{c}{84.03}  &  0.40 & \multicolumn{1}{c}{79.55} &  0.33 \\

100 & \multicolumn{1}{c}{230.23} &  1.30 & \multicolumn{1}{c}{136.60} &  0.37 & \multicolumn{1}{c}{124.19} &  0.24 & \multicolumn{1}{c}{118.92} &  0.19 \\

120 & \multicolumn{1}{c}{229.75} &  0.91 & \multicolumn{1}{c}{159.22} &  0.33 & \multicolumn{1}{c}{144.16} &  0.20 & \multicolumn{1}{c}{139.41} &  0.16 \\

140 & \multicolumn{1}{c}{249.38} &  0.78 & \multicolumn{1}{c}{173.49} & 0.24 & \multicolumn{1}{c}{163.90} &  0.17 & \multicolumn{1}{c}{158.90} &  0.14 \\
\bottomrule
\end{tabular}
\end{table*}

\subsection{Synthesis Quality}
\label{section:syntqual}
We use DNSMOS \cite{reddy2021dnsmos} as an objective measure of naturalness for our experiments. DNSMOS provides three ratings for quality of speech (SIG), noise (BAK), and overall (OVRL). Additionally, we assess the intelligibility of synthesized speech through Word Error Rate (WER). We calculated WER using an ASR consisting of a CRDNN based acoustic model and a transformer-based language model that uses CTC and attention decoders. Table \ref{tab:DNSMOS}, summarizes results for the five baselines and the proposed systems. In terms of DNSMOS, our models achieve comparable ratings as Diff-VC, QuickVC, and B3 across the three measures, and comparable or better than the source speech. In terms of intelligibility, our systems achieve comparable WER to that of B3, and superior to the rest, even though our models operate in a causal fashion with far more limited context.

We verified the synthesis quality of our two models through listening tests on Amazon Mechanical Turk (AMT).  Namely, participants (N=20) were asked to rate the acoustic quality of utterances using a standard 5-point scale mean opinion score (MOS) 
% (5: excellent, 1: bad).
[rating, speech quality, level of distortion]: [5, excellent, imperceptible] | [4, good, just perceptible but not annoying] | [3, fair, perceptible but slightly annoying] | [2, poor, annoying but not objectionable ] | [1, bad, very annoying and objectionable]. %labels are listed in Table \ref{tab:mosratings}. 
Each listener rated utterances synthesized using the base and lite models, as well as original utterances (20 for each). Results are shown in Table \ref{tab:MOS}. Both systems (base and lite) obtained comparable ratings of MOS as the original utterances. We see a difference of 0.1 between the MOS score of base and lite versions, but we didn't find them to be significant. It is noteworthy that while the lite version has 0.1x number of parameters, it achieves nearly the same synthesis quality as the base version.

\begin{table}[t]
  \caption{DNS MOS, Word Error Rates (WER) and speaker similarity scores (SSS) for baselines and the proposed model.}
  \label{tab:DNSMOS}
  \centering
  \begin{tabular}{llllll}
\toprule
& \textbf{SIG} & \textbf{BAK} & \textbf{OVRL} & \textbf{WER} & \textbf{SSS} \\ \midrule
VQMIVC    & 3.35  & 3.78 & 2.98 & 26.88 & 0.72 \\
Quick-VC  & 3.55  & 4.06 & 3.28 & 6.30  & 0.68 \\
Diff-VC   & 3.62  & 4.17 & 3.40 & 7.64  & 0.82 \\ 
VPC22 B2  & 2.85  & 3.44 & 2.50 & 12.02 & - \\
B3        & 3.57  & 4.03 & 3.28 & 4.65  & - \\
\midrule
base      & 3.53  & 3.99 & 3.31 & 5.12  & 0.85 \\ 
lite      & 3.48  & 3.93 & 3.22 & 6.47  & 0.81 \\ \midrule
Source    & 3.58  & 3.99 & 3.26 & 2.98 & 0.89 \\ \bottomrule
\end{tabular}
\end{table}

\begin{table}[t]
  \caption{Subjective MOS for naturalness.}
  \label{tab:MOS}
  \centering
  \begin{tabular}{llllll}
\toprule
& \textbf{Source} & \textbf{base} & \textbf{lite} \\ \midrule
MOS & 3.54 $\pm$ 0.56 & 3.57 $\pm$ 0.59   & 3.47 $\pm$ 0.62   \\
\bottomrule
\end{tabular}
\end{table}

\subsection{Speaker Anonymization}
To assess speaker-anonymization performance, we report Equal Error Rate (EER)  on the speaker verification model (ASV) in the VoicePrivacy 2024 Challenge github (see section \ref{sec:setup}). ASV tests are conducted for the following two scenarios, (a) ignorant, where we only anonymize the trial data (O-A), or (b) lazy-informed, where we anonymize both enrollment and trial data but use different target speakers (A-A). Results are shown in Table \ref{tab:EER}. For both the ignorant and lazy-informed scenario, our models achieves similar performance as B3 and outperforms VPC22 B2. Although our base model performs slightly worse than B3, the differences are not significantly different between them ($p=0.13$).
% Although B2 (VPC22) performs moderately well for the ignorant scenario, its performance drops significantly when the attacker is lazy-informed. For both the ignorant and lazy-informed scenario, B3 and our proposed models (ML based) outperform the DSP based method and achieve EER scores close to 45\%. Although EERs for B3 are slightly better than those of our systems, differences are not significantly different between them (p$\ll$ 0.001).

To corroborate these results, we conducted an AB listening test on AMT. Participants were presented with two audio samples, one from a speaker in the enrollment set, and the second sample from one of three options: (a) a different utterance from the same speaker from the trial set, (b) an utterance from a different speaker from the trial set, or (c) another utterance of the same speaker from the trial set but anonymized through our lite version of the system. Then, participants had to decide if both samples were from the same speaker, and rate the confidence in their decision using a 7-point scale (7: extremely confident; 5: quite a bit confident; 3: somewhat confident; 1: not confident at all). Each listener rated 20 AB pairs per scenario. Results are summarized in Table \ref{tab:SUBJECTIVE}.  In settings (a) and (b), listeners could easily identify whether the recording were from the same or different speakers (81 \% ) with high confidence (5.71). In setting (c), however, the anonymized trial data obtained obtained similar rating as in (b), indicating that proposed system was able to anonymize the trial recordings.

\begin{table}[t]
  \caption{Equal Error Rate (EER) as a privacy metric. The higher the better.}
  \label{tab:EER}
  \centering
  \begin{tabular}{llllll}
\toprule
& & \textbf{VPC22 B2}  & \textbf{B3} & \textbf{base} & \textbf{lite} \\ \midrule
O-A & M & 25.10   & 44.24 & 43.83  & 42.57   \\
    & F & 37.42   & 47.78 & 46.87  & 45.31   \\
\midrule
A-A & M & 11.03  & 42.63  & 41.43  & 39.20  \\ 
    & F & 15.03  & 43.23  & 42.03  & 41.16  \\
\bottomrule
\end{tabular}
\end{table}

\begin{table}[t]
  \caption{Subjective speaker verifiability scores for the proposed model.}
  \label{tab:SUBJECTIVE}
  \centering
  \begin{tabular}{llll}
\toprule
\textbf{speaker} & \textbf{anon} & \textbf{Verifiability} & \textbf{ Confidence Rating} \\ \midrule
same & no    & 81.5\% & 5.71   \\
different & no  & 17.75\% & 2.50   \\
same & yes   & 14.5\% & 2.37  \\
\bottomrule
\end{tabular}
\end{table}

\subsection{Speaker Identity transfer}
In a final step, we evaluated our models' ability to capture the voice of a target speaker.  For this purpose, we used an objective score of speaker similarity based on the cosine similarity between speaker embeddings of the target and the synthesized utterances obtained from a ASV system\footnote{Computed using: https://github.com/resemble-ai/Resemblyzer}. We compare our model against the three VC baselines (VQMIVC \cite{wang21n_interspeech}, QuickVC \cite{guo2023quickvc}, DiffVC \cite{popov2021diffusion}) using the same settings as those in section \ref{section:syntqual}) to generate VC samples. Results are summarized in the rightmost column of Table \ref{tab:DNSMOS} (SSS). As a guideline, pairing two utterances from the same speaker yields an average cosine similarity of 0.89. As shown, our base model outperforms the three baselines, achieving cosine similarity that is close to the average within-speaker similarity of 0.89.

\section{Discussion}
\label{sec:discussion}
Most existing speaker anonymization methods do not operate in low-latency streaming mode, preventing their use in field operations. In this paper, we present an end-to-end streaming model that operates with low latency and achieves anonymization by mapping speaker embedding into an artificially generated pseudo speaker in a causal fashion (i.e., no future context).
The pseudo-speaker generator can produce speaker embeddings that are very close to a real person in the corpus. Although we train the pseudo-speaker generator on a different corpus than the speech anonymization system to guard against this possibility, we could also test if a pseudo-speaker is too close to one on the corpus or outside the space of speakers in the training corpus, and generate new ones until a valid one is generated.
While there exists a quality-latency tradeoff, our system can achieve latency as low as $66 ms$ while maintaining state-of-the-art naturalness, intelligibility and privacy preservation. Our lite version is roughly 10MB and can potentially be deployed on mobile devices to support real-time field applications. Accent can carry speaker related cues \cite{das2020understanding} and in future work, we aim to add accent conversion to this pipeline. Other research direction is to add the control of emotion while synthesizing speech signals.

\section{Acknowledgements}
This work was funded by NSF award 619212 and 1623750.

\bibliographystyle{IEEEbib}
\bibliography{strings,refs}

\end{document}